\documentclass[10pt,conference]{IEEEtran}

\newif\iftwocols

\usepackage{amsmath,amsthm,algorithm}
\usepackage{tikz,tkz-graph}

\newcommand{\ve}{\bf}

\usepackage{graphicx}
\usepackage{color}
\usepackage{epsfig, amsmath, amsthm, amssymb,latexsym,amsmath,amsbsy, rotating}
\usepackage{amsfonts}

\newtheorem{thm}{Theorem}
\newtheorem{lemma}{Lemma}

\newtheorem{remark}{Remark}

\newtheorem{algo}{Algorithm}

\theoremstyle{definition}

\newtheorem{defn}{Definition}
\DeclareFontFamily{U}{futm}{}
\DeclareFontShape{U}{futm}{m}{n}{
  <-> s * [.92] fourier-bb
  }{}
\DeclareSymbolFont{Ufutm}{U}{futm}{m}{n}
\DeclareSymbolFontAlphabet{\mathbb}{Ufutm}

\begin{document}

\title{Common Information Components Analysis}
\author{
\IEEEauthorblockN{Michael Gastpar and Erixhen Sula} \\
\IEEEauthorblockA{School of Computer and Communication Sciences\\EPFL\\ Lausanne,
Switzerland\\
\{michael.gastpar, erixhen.sula\}@epfl.ch}
}

\maketitle

\begin{abstract}
We give an information-theoretic interpretation of Canonical Correlation Analysis (CCA) via (relaxed) Wyner's common information. CCA permits to extract from two high-dimensional data sets low-dimensional descriptions (features) that capture the commonalities between the data sets, using a framework of correlations and linear transforms. Our interpretation first extracts the common information up to a pre-selected resolution level, and then projects this back onto each of the data sets. In the case of Gaussian statistics, this procedure precisely reduces to CCA, where the resolution level specifies the number of CCA components that are extracted. This also suggests a novel algorithm, Common Information Components Analysis (CICA), with several desirable features, including a natural extension to beyond just two data sets.

\end{abstract}


\IEEEpeerreviewmaketitle

\section{Introduction}

Understanding relations between two (or more) sets of variates is key to many tasks in data analysis and beyond.
To approach this problem, it is natural to reduce each of the sets of variates separately in such a way that the reduced descriptions, or {\it features,} fully capture the {\it commonality}  between the two sets,
while suppressing aspects that are individual to each of the sets.
This permits to understand the relation between the data sets without obfuscation.

A popular framework to accomplish this task follows the classical viewpoint of {\it dimensionality reduction } and is referred to as Canonical Correlation Analysis (CCA)~\cite{Hotelling:36}.
CCA seeks the best {\it linear} extraction, i.e., we consider linear projections of the original variates.
In this case, the quality of the extraction is assessed via the resulting correlation coefficient.
The result can be expressed directly via the singular value decomposition.
Via the so-called Kernel trick, this can be extended to cover arbitrary (fixed) function classes.

An alternative framework is built around the concept of maximal correlation. Here, one seeks arbitrary (not necessarily linear) remappings of the original data in such a way as to maximize their correlation coefficient. This perspective culminates in the well-known {\it alternating conditional expectation } (ACE) algorithm~\cite{BreimanF:85},
but the problem does not admit a compact solution.

In both approaches, the commonality between variates is measured by correlation.
By contrast, in this paper, we consider a different approach that measures commonality between variates via (relaxed Wyner's) Common Information~\cite{GastparS:19itw,SulaG:19it}, a variant of a mutual information measure.

\subsection{Contributions}

\begin{figure}
\begin{center}
\begin{tikzpicture}
\draw[black,thick] (0,5.7) node[anchor=west]{Original Data};
\draw[black,thick] (0,5.3) node[anchor=west]{Sets};
\draw[black,thick] (4,5.5) node{${\bf X}_1, \ldots, {\bf X}_n$};
\draw[black,thick] (7,5.5) node{${\bf Y}_1, \ldots, {\bf Y}_n$};

\draw[->,black,thick](4.2,5.2)--(5.3,3.8);
\draw[->,black,thick](6.8,5.2)--(5.7,3.8);

\draw[black,thick] (0,3.7) node[anchor=west]{Common Information};
\draw[black,thick] (0,3.3) node[anchor=west]{at level $\gamma$};
\draw[black,thick] (5.5,3.5) node{${\bf W}_1, \ldots, {\bf W}_n$};

\draw[->,black,thick](5.3,3.2)--(4.2,1.8);
\draw[->,black,thick](5.7,3.2)--(6.8,1.8);

\draw[black,thick] (0,1.9) node[anchor=west]{Common Information};
\draw[black,thick] (0,1.5) node[anchor=west]{Components};
\draw[black,thick] (4,1.5) node{${\bf U}_1, \ldots, {\bf U}_n$};
\draw[black,thick] (7,1.5) node{${\bf V}_1, \ldots, {\bf V}_n$};

\draw[black,thick] (5.5,2.3) node{projection};

\end{tikzpicture}
\caption{Common Information Components Analysis (for the case of two data sets): For two (high-dimensional) data sets (sources) $\bf X$ and $\bf Y,$ we first determine their (Wyner's) common information. The Common Information Components are then obtained by projecting the common information back onto the two data sources, respectively. The parameter $\gamma$ is the compression level: a larger $\gamma$ means coarser common information.}\label{fig-concept}
\end{center}
\end{figure}
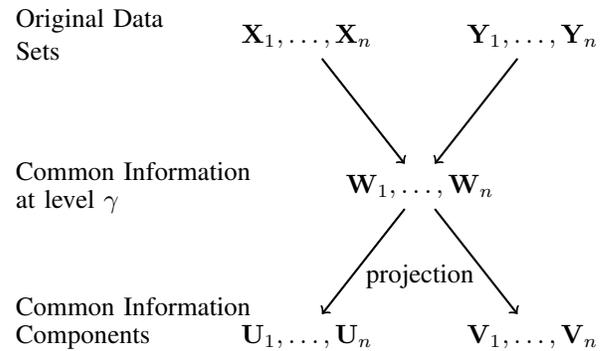

The main contributions of our work are:
\begin{itemize}
\item The introduction of a novel algorithm, referred to as Common Information Components Analysis (CICA), to separately reduce each set of variates in such a way as to retain the commonalities between the sets of variates while suppressing their individual features. A conceptual sketch is given in Figure~\ref{fig-concept}.
\item The proof that for the special case of Gaussian variates, CICA reduces to CCA. Thus, CICA is a strict generalization of CCA.
\end{itemize}

\subsection{Related Work}

Connections between CCA and Wyner's common information have been explored in the past.
It is well known that for Gaussian vectors, (standard, non-relaxed) Wyner's common information is attained by all of the CCA components together, see~\cite{SatpathyC:15}.
This has been further interpreted, see e.g.~\cite{HuangWZ:18itw}.
To put our work into context, we note it is only the {\it relaxed}  Wyner's common information~\cite{GastparS:19itw,SulaG:19it} that permits to conceptualize the sequential, one-by-one recovery of the CCA components,
and thus, the spirit of dimensionality reduction.

Information measures have played a role in earlier considerations with some connections to dimensionality reduction and feature extraction.
This includes independent components analysis (ICA)~\cite{Comon:91} and the information bottleneck~\cite{WitsenhausenW:75,TishbyPB:99}, amongst others.
Finally, we note that an interpretation of CCA as a (Gaussian) probabilistic model was presented in~\cite{BachJ:05}.

\subsection{Notation}

A bold capital letter such as ${\bf X}$ denotes a random vector, and ${\bf x}$ its realization.
A non-bold capital letter such as $K$ denotes a (fixed) matrix, and $K^H$ its Hermitian transpose. Specifically,
$K_{\bf X}$ denotes the covariance matrix of the random vector $\bf X.$
$K_{\bf X Y}$ denotes the covariance matrix between random vectors ${\bf X}$ and ${\bf Y.}$

\section{Relaxed Wyner's Common Information}

The main framework and underpinning of the proposed algorithm is Wyner's common information and its extension, which is briefly reviewed in the sequel,
along with its key properties.

\subsection{Wyner's Common Information}

Wyner's common information is defined for two random variables (or random vectors) $X$ and $Y$ of arbitrary fixed joint distribution $p(x,y).$

\begin{defn}[from~\cite{Wyner:75}]
For random variables $X$ and $Y$ with joint distribution $p(x,y),$ Wyner's common information is defined as
\begin{align}
C(X; Y) &=  \min_{p(w|x,y)} I(X,Y ; W) \mbox{ such that } I(X;Y|W) =0.
\end{align}
\end{defn}

Basic properties are stated below in Lemma~\ref{Lemma-relWyner-basicprops} (setting $\gamma=0$).
We note that explicit formulas for Wyner's common information are known only for a small number of special cases.
The case of the doubly symmetric binary source is solved completely in~\cite{Wyner:75} and can be written as
\begin{align}
 C(X;Y) &= 1 + h_b(a_0) - 2 h_b\left(\frac{1-\sqrt{1-2a_0}}{2}\right),    \label{Eq-WCI-binary-WynerParameterization}
\end{align}
where $a_0$ denotes the probability that the two sources are unequal (assuming without loss of generality $a_0\le \frac{1}{2}$).
Further special cases of discrete-alphabet sources appear in~\cite{Witsenhausen:76}. 

Moreover, when $X$ and $Y$ are jointly Gaussian with correlation coefficient $\rho,$ then
$C(X;Y) = \frac{1}{2} \log \frac{1 + |\rho|}{1-|\rho|}.$
Note that for this example, $I(X;Y) = \frac{1}{2} \log \frac{1}{1-\rho^2}.$
This case was solved in~\cite{Xu--Liu--Chen,Xu--Liu--Chen-2} using a parameterization of conditionally independent distributions.
We note that an alternative proof follows from the arguments presented in~\cite{GastparS:19itw,SulaG:19it}.

\subsection{Relaxed Wyner's Common Information}

\begin{defn}[from~\cite{GastparS:19itw}]\label{def-Wyner-relaxed}
For random variables $X$ and $Y$ with joint distribution $p(x,y),$ the relaxed Wyner's common information is defined as (for $\gamma \ge 0$)
\begin{align}
C_{\gamma} (X; Y) &=  \min_{p(w|x,y)} I(X,Y ; W) \mbox{ such that } I(X;Y|W) \le \gamma. \label{Eq-def-Wyner-relaxed}
\end{align}
\end{defn}

\begin{lemma}[from~\cite{SulaG:19it}]\label{Lemma-relWyner-basicprops}
The relaxed Wyner's common information satisfies the following properties:
\begin{enumerate}
\item For discrete $X$ and $Y,$ the cardinality of ${\cal W}$ may be restricted to $|{\cal W}| \le |{\cal X}| |{\cal Y}| + 1.$
\item $C_\gamma(X;Y) \ge 0$ with equality if and only if $\gamma \ge I(X;Y).$
\item $C_{\gamma} (X; Y) \ge \max\{ I(X;Y)-\gamma, 0\}.$
\item Data processing inequality: If $X-Y-Z$ form a Markov chain, then $C_\gamma(X;Z)\le \min\{ C_\gamma(X;Y), C_\gamma(Y;Z)\}.$
\item $C_\gamma(X;Y)$ is a convex and continuous function of $\gamma$ for $\gamma \ge 0.$
\item If $f(\cdot)$ and $g(\cdot)$ are one-to-one functions, then $C_{\gamma}(f(X); g(Y)) = C_\gamma(X;Y).$
\item For discrete $X,$ we have $C_\gamma(X;X) = \max\{H(X)-\gamma,0\}.$
\item Let $\{ (X_i,Y_i) \}_{i=1}^n$ be $n$ independent pairs of random variables. Then
\begin{align}
C_{\gamma} (X^n; Y^n) &=  \min_{\{\gamma_i\}_{i=1}^n : \sum_{i=1}^n \gamma_i=\gamma}  \sum_{i=1}^n C_{\gamma_i}(X_i; Y_i).  \label{Eqn-thm:gensplit}
\end{align}
\end{enumerate}
\end{lemma}

Explicit formulas for the relaxed Wyner's common information are not currently known for most $p(x,y).$
A notable exception is when $\ve X$ and $\ve Y$ are jointly Gaussian random vectors of length $n.$
Denote the covariance matrices of the vectors $\ve X$ and $\ve Y$ by $K_{\ve X}$ and $K_{\ve Y},$ respectively,
and the covariance matrix between $\ve X$ and $\ve Y$ by $K_{\ve X \ve Y}.$
Then (see~\cite{SulaG:19it}),
\begin{align}
&C_{\gamma}({\ve X};{\ve Y})= \min_{\gamma_i: \sum_{i=1}^n \gamma_i = \gamma} \sum_{i=1}^n C_{\gamma_i}(X_i;Y_i), \label{eq-relwyner-gaussvec-1}
\end{align}
where 
\begin{align}
C_{\gamma_i}(X_i;Y_i)=\frac{1}{2} \log^+ \frac{(1+\rho_i)(1-\sqrt{1- e^{-2 \gamma_i}})}{(1-\rho_i)(1+\sqrt{1- e^{-2 \gamma_i}})}\label{eq-relwyner-gaussvec-2}
\end{align}
and $\rho_i$ (for $i=1,\dots,n$) are the singular values of $K_{\ve X}^{-1/2} K_{\ve X \ve Y} K_{\ve Y}^{-1/2}$.
By contrast, for the doubly symmetric binary source, the relaxed Wyner's common information is currently unknown
(a bound and conjecture appear in~\cite{SulaG:19it}).

%

\section{The Algorithm}

In this section, we present the proposed algorithm in the idealized setting of unlimited data.
Specifically, for the proposed algorithm, this means that we assume perfect knowledge of the data distribution $p({\bf x}, {\bf y}).$

\subsection{High-level Description}

The idea of the proposed algorithm is to estimate the relaxed Wyner's Common Information of Equation~\eqref{Eq-def-Wyner-relaxed} between the information sources (data sets) at the chosen level $\gamma.$
This estimate will come with an associated conditional distribution $p_{\gamma}(w|x,y).$
Obtaining the dimension-reduced versions then can be thought of as a type of projection of the resulting random variable $W$ back on $X$ and $Y,$ respectively.
For the case of Gaussian statistics, this can be made precise.

\subsection{Main Steps of the Algorithm}

The algorithm proposed here starts from the joint distribution of the data, $p({\bf x}, {\bf y}).$ Estimates of this distribution can be obtained from data samples ${\bf X}^n$ and ${\bf Y}^n$
via standard techniques.
The main steps of the procedure can then be described as follows:

\begin{algo}[CICA]\label{alg-CICA-generic}
\begin{enumerate}
\item Select a real number $\gamma,$ where $0 \le \gamma \le I({\bf X}; {\bf Y}).$
This is the compression level: A low value of $\gamma$ represents low compression, and thus, many components are retained.
A high value of $\gamma$ represents high compression, and thus, only a small number of components are retained.
\item Solve the relaxed Wyner's common information problem,
\begin{align}
\min_{p(w|{\bf x,y})} I({\bf X, Y} ; W) \mbox{ such that } I({\bf X};{\bf Y}|W) \le \gamma,
\end{align}
leading to an associated conditional distribution $p_{\gamma}(w|{\bf x}, {\bf y}).$\footnote{We note that this is not generally unique. For example, if $W$ is a minimizer, then so is $g(W)$ for any one-to-one mapping $g(\cdot).$}

\item The dimension-reduced data sets are

\begin{enumerate}
\item Version 1: MAP (maximum {\it a posteriori}):
\begin{itemize}
\item $u({\bf x}) = \arg \max_{w} p_{\gamma}(w|{\bf x})$
\item $v({\bf y}) =\arg \max_{w} p_{\gamma}(w|{\bf y})$
\end{itemize}
\item Version 2: Conditional Expectation:
\begin{itemize}
\item $u({\bf x}) = {\mathbb E}[W | {\bf X} = {\bf x}]$
\item $v({\bf y}) = {\mathbb E}[W | {\bf Y} = {\bf y}]$
\end{itemize}
\item Version 3: Marginal Integration:
\begin{itemize}
\item $u({\bf x}) = \int_{\bf y} p({\bf y}) {\mathbb E}[ W | {\bf X}={\bf x}, {\bf Y}={\bf y} ] d{\bf y}$
\item $v({\bf y}) =\int_{\bf x} p({\bf x}) {\mathbb E}[ W | {\bf X}={\bf x}, {\bf Y}={\bf y} ] d{\bf x}$
\end{itemize}
\end{enumerate}
\end{enumerate}
\end{algo}

\subsection{A binary toy example}\label{Sec-Ex-binary}

Let us illustrate the proposed algorithm via a simple toy example.
Consider the vector $(\tilde{X}_1, \tilde{X}_2, \tilde{Y}_1, \tilde{Y}_2)$ of binary random variables.
Suppose that $(\tilde{X}_1,  \tilde{Y}_1)$ are a doubly symmetric binary source ({\it i.e., } $\tilde{X}_1$ is uniform, and $\tilde{X}_2$  is the result of passing $\tilde{X}_1$ through a binary symmetric (``bit-flipping'') channel)
while $\tilde{X}_2$ and $\tilde{Y}_2$ are independent binary uniform random variables
(also independent of $(\tilde{X}_1,  \tilde{Y}_1)$).
We will then form the vectors ${\bf X}$ and ${\bf Y}$ as
\begin{align}
{\bf X} &= \left( \begin{array}{c}  \tilde{X}_1 \oplus \tilde{X}_2 \\
                                                    \tilde{X}_2
                         \end{array} \right),
\end{align}
and
\begin{align}
{\bf Y} &= \left( \begin{array}{c}  \tilde{Y}_1 \oplus \tilde{Y}_2 \\
                                                     \tilde{Y}_2
                         \end{array} \right),
\end{align}
where $\oplus$ denotes the modulo-reduced addition, as usual.
Observe that any pair amongst the four entries in these two vectors are (pairwise) independent binary uniform random variables.
Hence, the overall covariance matrix of the merged random vector $({\bf X}^T, {\bf Y}^T)^T$ is merely a scaled identity matrix, implying that CCA does not do anything.

By contrast, for the CICA algorithm (with $\gamma=0$ and using the MAP version), an optimal solution is to reduce ${\bf X}$ to $\tilde{X}_1$ and ${\bf Y}$ to $\tilde{Y}_1.$ This captures all the dependence between the vectors ${\bf X}$ and ${\bf Y},$ which appears to be the most desirable outcome.

\section{For Gaussian, CICA is CCA}

In this section, we consider the proposed CICA algorithm in the idealized setting where the data distribution $p({\bf x}, {\bf y})$ is known exactly.
Specifically, we establish that if $p({\bf x}, {\bf y})$ is a (multivariate) Gaussian distribution,
then the classic CCA is a solution to all versions of the proposed CICA algorithm.
This is the main technical contribution of the present work.

CCA is perhaps best described by first changing coordinates,
\begin{align}
\hat{\bf X} &= K_{\bf X}^{-1/2}{\bf X}  \\
\hat{\bf Y} &= K_{\bf Y}^{-1/2}{\bf Y} .
\end{align}
With this, the covariance matrix of the vector $\hat{\bf X}$ is the identity matrix,
and so is the covariance matrix of the vector $\hat{\bf Y}.$
CCA is then easily described by considering the covariance matrix between these two vectors,
\begin{align}
K_{\hat{\bf X}\hat{\bf Y}} & = K_{\bf X}^{-1/2} K_{\bf XY}K_{\bf Y}^{-1/2}.
\end{align}
A brief overview is given in Appendix~\ref{app-CCA}.
Let us denote the singular value decomposition of this matrix by
\begin{align}
K_{\hat{\bf X}\hat{\bf Y}} & = U \Sigma V^H,
\end{align}
where $\Sigma$ contains, on its diagonal, the ordered singular values of this matrix, denoted by $\rho_1 \ge \rho_2 \ge \ldots \ge \rho_n.$
CCA then performs the dimensonality reduction
\begin{align}
u({\bf x}) &= U_k^H\hat{\bf x} = U_k^HK_{\bf X}^{-1/2}{\bf x} \label{Eq-CCA-topkX} \\
v({\bf y}) &= V_k^H\hat{\bf y} = V_k^HK_{\bf Y}^{-1/2}{\bf y}, \label{Eq-CCA-topkY}
\end{align}
where the matrix $U_k$ contains the first $k$ columns of $U$ (that is, the $k$ left singular vectors corresponding to the largest singular values),
and the matrix $V_k$ the respective right singular vectors.
We refer to these as the ``top $k$ CCA components.''

\begin{thm}\label{thm-Gauss}
Let ${\bf X}$ and ${\bf Y}$ be jointly Gaussian random vectors. 
Then, the top $k$ CCA components are a solution to {\bf all three}  versions of Algorithm~\ref{alg-CICA-generic},
and $\gamma$ controls the number $k$ as follows:
\begin{align}
 k &= \left\{ \begin{array}{ll}
              n, & \mbox{ if }  0 \le \gamma < n I(\rho_n), \\
              n-1, & \mbox{ if }  n I(\rho_n) \le \gamma < (n-1) I(\rho_{n-1}) + I(\rho_n), \\
              n-2, & \mbox{ if }  (n-1) I(\rho_{n-1}) + I(\rho_n) \le \gamma \\
                &  \,\,\,\,\,\,\,\,  < (n-2) I(\rho_{n-2}) + I(\rho_{n-1}) + I(\rho_n), \\
              \vdots, & \vdots, \\
              \ell & \mbox{ if }  (\ell+1) I(\rho_{\ell+1}) + \sum_{i=\ell+2}^n I(\rho_i) \le \gamma \\
                &  \,\,\,\,\,\,\,\,  < \ell I(\rho_{\ell}) + \sum_{i=\ell+1}^n I(\rho_i), \\
                \vdots, & \vdots, \\
               1, & \mbox{ if } 2 I(\rho_2) + \sum_{i=2}^n I(\rho_i) \le \gamma <  \sum_{i=1}^n I(\rho_i), \\
              0, & \mbox{ if } \sum_{i=1}^n I(\rho_i) \le \gamma,
 \end{array} \right.
\end{align}
where $I(\rho) = \frac{1}{2} \log \frac{1}{1-\rho^2}.$
\end{thm}

\begin{remark}
Note that $k(\gamma)$ is a decreasing, integer-valued function.
\end{remark}

This theorem is a consequence of the main result in~\cite{GastparS:19itw}.
A proof outline is provided in Appendix~\ref{app-thm-Gauss}.

As mentioned earlier, the connection between CCA and (standard non-relaxed) Gaussian Wyner's common information is well known~\cite{SatpathyC:15}.
What is new in the present paper is the extension of this insight to {\it relaxed}  Wyner's common information.
This extension permits to extract the CCA components one-by-one via the compression parameter $\gamma.$
Evidently, the CICA algorithm only makes sense because we can tune how much common information we wish to extract.
In this sense, the choice $\gamma=0$ (the non-relaxed case) is not interesting since it amounts to a one-to-one transform of the original data (up to completely independent portions),
and thus, fails to capture the spirit of ``dimensionality reduction.''

\section{Extension to More Than Two Sources}

It is unclear how one would extend CCA to more than two databases.
By contrast, for CICA, this extension is conceptually straightforward.
The definition of relaxed Wyner's common information is readily extended to the general case:

\begin{defn}[Relaxed Wyner's Common Information for $M$ variables]\label{def-relaxed-Wyner-multiple}
For a fixed probability distribution $p(x_1, x_2, \ldots, x_M),$ we define
\begin{align}
C_\gamma(X_1; X_2; \ldots; X_M) &= \min I(X_1, X_2, \ldots, X_M ; W)
\end{align}
such that $\sum_{i=1}^M H(X_i|W) - H(X_1, X_2, \ldots, X_M|W)\le\gamma,$
where the minimum is over all probability distributions $p(w, x_1, x_2, \ldots, x_M)$ with marginal $p(x_1, x_2, \ldots, x_M).$
\end{defn}

Hence, to extend CICA (Algorithm~\ref{alg-CICA-generic}) to the case of $M$ databases, it now suffices to replace Step 2) with Definition~\ref{def-relaxed-Wyner-multiple}.
In Step 3), for all three versions, it is immediately clear how they can be extended. For example, for Version 1), we use
\begin{align}
u_i({\bf x}_i) & = \arg \max_{w} p_{\gamma}(w|{\bf x}_i),
\end{align}
for $i=1, 2, \ldots, M.$

It will be shown elsewhere how one can obtain the analogs of Equations~\eqref{eq-relwyner-gaussvec-1}-\eqref{eq-relwyner-gaussvec-2} for this generalized case, and thus, an extended version of Theorem~\ref{thm-Gauss}.

\section{Concluding Remarks and Future Work}

In a practical setting, one does not have access to the correct data distribution $p({\bf x}, {\bf y}).$
A first version is to simply work with an estimate of this distribution, based on the data available.
But a more interesting implementation is to combine the estimation step with the optimization step.
A fast algorithmic implementation will be presented elsewhere.

%
%

\appendices

\section{CCA}\label{app-CCA}

A brief review of CCA~\cite{Hotelling:36} is presented, mostly in view of the proof of Theorem~\ref{thm-Gauss}, given below in Appendix~\ref{app-thm-Gauss}.
Let ${\bf X}$ and ${\bf Y}$ be zero-mean real-valued random vectors with covariance matrices $K_{\bf X}$ and $K_{\bf Y},$ respectively. Moreover, let $K_{\bf XY}  =  { \mathbb E}[ {\bf X} {\bf Y}^H ].$
Let us first form
\begin{align}
\hat{\bf X} &= K_{\bf X}^{-1/2}{\bf X}, \label{Eq-CCA-coordX} \\
\hat{\bf Y} &= K_{\bf Y}^{-1/2}{\bf Y} .\label{Eq-CCA-coordY}
\end{align}
With this, the covariance matrix of the vector $\hat{\bf X}$ is the identity matrix,
and so is the covariance matrix of the vector $\hat{\bf Y}.$
CCA seeks to find vectors ${\bf u}$ and ${\bf v}$ such as to maximize the correlation between ${\bf u}^H\hat{\bf X}$ and ${\bf v}^H \hat{\bf Y},$ that is,
\begin{align}
  \max_{{\bf u}, {\bf v}} \frac{ { \mathbb E}[ {\bf u}^H\hat{\bf X}  \hat{\bf Y} ^H {\bf v} ] }{ \sqrt{{\mathbb E}[|{\bf u}^H\hat{\bf X} |^2]}\sqrt{{\mathbb E}[|{\bf v}^H\hat{\bf Y} |^2]}}, \label{Eq-CCA-problem}
\end{align}
which can be rewritten as
\begin{align}
  \max_{{\bf u}, {\bf v}} \frac{ {\bf u}^H K_{\bf \hat X \hat Y} {\bf v} }{ \|{\bf u} \| \,\, \| {\bf v} \|} , 
\end{align}
where
\begin{align}
K_{\bf \hat X \hat Y} &=   K_{\bf X}^{-1/2} K_{\bf XY}K_{\bf Y}^{-1/2} .
\end{align}
Note that this expression is {\it invariant}  to arbitrary (separate) scaling of ${\bf u}$ and ${\bf v}.$
To obtain a unique solution, we could choose to impose that both vectors be unit vectors,
\begin{align}
 \max_{{\bf u}, {\bf v}: \|{\bf u} \| = \| {\bf v} \| =1}  {\bf u}^H K_{\bf \hat X \hat Y} {\bf v} .
\end{align}
From Cauchy-Schwarz, for a fixed ${\bf u},$ the maximizing (unit-norm) ${\bf v}$ is given by
\begin{align}
  {\bf v} &=  \frac{K_{\bf \hat X \hat Y}^H  {\bf u}}{ \left\| K_{\bf \hat X \hat Y}^H  {\bf u} \right\| } ,\label{Eqn-CCA-relationbetweenvectors}
\end{align}
or equivalently, for a fixed ${\bf v},$ the maximizing (unit-norm) ${\bf u}$ is given by
\begin{align}
  {\bf u} &=  \frac{K_{\bf \hat X \hat Y}  {\bf v}}{ \left\| K_{\bf \hat X \hat Y}  {\bf v} \right\| } .
\end{align}
Plugging in the latter, we obtain
\begin{align}
 \max_{{\bf v}: \|{\bf v} \| = 1}  \frac{ {\bf v}^H K_{\bf \hat X \hat Y}^H K_{\bf \hat X \hat Y} {\bf v} }{ \left\| K_{\bf \hat X \hat Y}  {\bf v} \right\| } ,
\end{align}
or, dividing through,
\begin{align}
 \max_{{\bf v}: \|{\bf v} \| = 1}   \left\| K_{\bf \hat X \hat Y} {\bf v} \right\| .
 \end{align}
 The solution to this problem is well known:
${\bf v}$ is the right singular vector corresponding to the largest singular vector of the matrix $K_{\bf \hat X \hat Y}=K_{\bf X}^{-1/2} K_{\bf XY}K_{\bf Y}^{-1/2}.$
Evidently, ${\bf u}$ is the corresponding left singular vector.
Restarting again from Equation~\eqref{Eq-CCA-problem}, but restricting to vectors that are orthogonal to the optimal choices of the first round leads to the second CCA components, and so on.

\section{Proof Outline for Theorem~\ref{thm-Gauss}}\label{app-thm-Gauss}

In the case of Gaussian vectors, the solution to the optimization problem in Equation~\eqref{Eq-def-Wyner-relaxed} is most easily described in two steps.
First, we apply the change of basis indicated in Equations~\eqref{Eq-CCA-coordX}-\eqref{Eq-CCA-coordY}.
This is a one-to-one transform, leaving all information expressions in Equation~\eqref{Eq-def-Wyner-relaxed} unchanged.
In the new basis, we have $n$ independent pairs.
When $X$ and $Y$ consist of independent pairs, the solution to the optimization problem in Equation~\eqref{Eq-def-Wyner-relaxed}
can be reduced to $n$ separate scalar optimizations, see~\cite[Theorem 3]{SulaG:19it} (also quoted above in Lemma~\ref{Lemma-relWyner-basicprops}, Item 8).
The remaining crux then is solving the scalar Gaussian version of the optimization problem in Equation~\eqref{Eq-def-Wyner-relaxed}.
This is done in~\cite[Theorem 4]{SulaG:19it} via an argument of factorization of convex envelope.
The full solution to the optimization problem is given in Equation~\eqref{eq-relwyner-gaussvec-1}-\eqref{eq-relwyner-gaussvec-2}.
The remaining allocation problem over the non-negative numbers $\gamma_i$ can be shown to lead to a water-filling solution, see~\cite[Section IV]{SulaG:19it}.
More explicitly, to understand this solution, start by setting $\gamma=I({\bf X}; {\bf Y}).$ Then, the corresponding $C_{\gamma}({\bf X}; {\bf Y})=0$
and the optimizing distribution $p_{\gamma}(w|{\bf x}, {\bf y})$ trivializes.
Now, as we lower $\gamma,$ the various terms in the sum in Equation~\eqref{eq-relwyner-gaussvec-1} start to become non-zero, starting with the term with the largest correlation coefficient $\rho_1.$
Hence, an optimizing distribution $p_{\gamma}(w|{\bf x}, {\bf y})$ can be expressed as ${\bf W}_{\gamma} = U_k^HK_{\bf X}^{-1/2}{\bf X} + V_k^HK_{\bf Y}^{-1/2}{\bf Y} + {\bf Z},$
where the matrices $U_k$ and $V_k$ are precisely the top $k$ CCA components (see Equations~\eqref{Eq-CCA-topkX}-\eqref{Eq-CCA-topkY} and the following discussion),
and ${\bf Z}$ is additive Gaussian noise with mean zero, independent of ${\bf X}$ and ${\bf Y}.$

For the algorithm, we need the corresponding conditional marginals, $p_{\gamma}(w|{\bf x})$ and $p_{\gamma}(w|{\bf y}).$
By symmetry, it suffices to prove one formula. Changing basis as in Equations~\eqref{Eq-CCA-coordX}-\eqref{Eq-CCA-coordY}, we can write
\begin{align}
{\mathbb E}[ W | {\bf X} ] &=  {\mathbb E}[ U_k^H\hat{\bf X} + V_k^H\hat{\bf Y} + {\bf Z} | \hat{\bf X} ] \\
  &= U_k^H \hat{\bf X} +  V_k^H{\mathbb E}[ \hat{\bf Y} | \hat{\bf X} ]  \\
  &= U_k^H \hat{\bf X} + V_k^H \left(  {\mathbb E}[ \hat{\bf Y}  \hat{\bf X}^H ] \left(  {\mathbb E}[ \hat{\bf X}  \hat{\bf X}^H ] \right)^{-1} \hat{\bf X} \right)   \\
   & = U_k^H \hat{\bf X} +V_k^HK_{\hat{\bf Y}\hat{\bf X}}\hat{\bf X} \\
   & =  U_k^H \hat{\bf X} + \left(  K_{\hat{\bf X}\hat{\bf Y}}^HV_k\right)^H \hat{\bf X}  .
\end{align}
Finally, note that Equation~\eqref{Eqn-CCA-relationbetweenvectors} can be read as
\begin{align}
  {\bf u} &=  \alpha  K_{\hat{\bf X}\hat{\bf Y}}^H {\bf v},
\end{align}
for some real-valued constant $\alpha.$
Thus, combining the top $k$ CCA components,
\begin{align}
  U_k &=  D  K_{\hat{\bf X}\hat{\bf Y}}^H V_k,
\end{align}
where $D$ is a diagonal matrix.
Hence,
 \begin{align}
{\mathbb E}[ W | {\bf X} ] &= U_k^H \hat{\bf X} + D^{-1} U_k^H  \hat{\bf X} \\
 &= \tilde{D} U_k^H \hat{\bf X} , \label{Eq-proof-EWX-final}
\end{align}
where $\tilde{D}$ is the diagonal matrix
 \begin{align}
\tilde{D} &= I + D^{-1} .
\end{align}
This is precisely the top $k$ CCA components (note that the solution to the CCA problem~\eqref{Eq-CCA-problem} is only specified up to a scaling).
This establishes the theorem for the case of Version 2) of the proposed algorithm.
Clearly, it also establishes that $p_{\gamma}(w|{\bf x})$ is a Gaussian distribution with mean given by~\eqref{Eq-proof-EWX-final}, thus establishing the theorem for Version 1) of the proposed algorithm.
The proof for Version 3) follows along similar lines and is thus omitted.

\section*{Acknowledgment}
This work was supported in part by the Swiss National Science Foundation under Grant 169294, Grant P2ELP2\_165137. 
\bibliographystyle{IEEEtran}
\bibliography{CICAbib}

\end{document}